\documentclass[11pt,a4paper]{article}
\pdfoutput=1

\usepackage[vmargin=25mm,hmargin=25mm]{geometry}
\usepackage{amsmath,amssymb,amsthm,booktabs,color,enumitem,graphicx,latexsym,url,xspace}
\usepackage[numbers,sort&compress]{natbib}
\usepackage[basic,small]{complexity}
\usepackage[small]{caption}
\usepackage[outercaption]{sidecap}
\usepackage{microtype}
\usepackage{hyperref}

\newtheorem{theorem}{Theorem}

\newtheorem{lemma}[theorem]{Lemma}

\theoremstyle{definition}

\newclass{\LOCAL}{LOCAL}
\newcommand{\A}{\mathcal{A}}

\hypersetup{
    colorlinks=true,
    linkcolor=black,
    citecolor=black,
    filecolor=black,
    urlcolor=[rgb]{0,0.1,0.5},
    pdfauthor={Miikka Hilke, Christoph Lenzen, Jukka Suomela},
    pdftitle={Local approximability of minimum dominating set on planar graphs},
}

\begin{document}

\begin{flushleft}
{\Large\bfseries Local Approximability of \\ Minimum Dominating Set on Planar Graphs\par}
\bigskip
\bigskip

\textbf{Miikka Hilke}
$\cdot$
{\sffamily\small miikka.hilke@helsinki.fi}
$\cdot$
{\footnotesize
Helsinki Institute for Information Technology HIIT,
Department of Computer Science, University of Helsinki, Finland\par}

\medskip
\textbf{Christoph Lenzen}
$\cdot$
{\sffamily\small clenzen@csail.mit.edu}
$\cdot$
{\footnotesize
Computer Science and Artificial Intelligence Laboratory, MIT\par}

\medskip
\textbf{Jukka Suomela}
$\cdot$
{\sffamily\small jukka.suomela@aalto.fi}
$\cdot$
{\footnotesize
Helsinki Institute for Information Technology HIIT,
Department of Information and Computer Science, Aalto University\par}
\end{flushleft}

\bigskip
\noindent\textbf{Abstract.}
We show that there is no deterministic local algorithm (constant-time distributed graph algorithm) that finds a $(7-\epsilon)$-approximation of a minimum dominating set on planar graphs, for any positive constant $\epsilon$. In prior work, the best lower bound on the approximation ratio has been $5-\epsilon$; there is also an upper bound of $52$.

\medskip

\section{Introduction}

This work studies one of the last uncharted corners in the area of deterministic local algorithms: planar graphs.

A \emph{local algorithm} is a distributed graph algorithm that runs in $O(1)$ communication rounds, independently of the size of the network. While the theory of \emph{randomised} local algorithms is still in its infancy, we have nowadays a good understanding of the capabilities of \emph{deterministic} local algorithms.

For many classical graph problems, there are exactly matching upper and lower bounds on the best possible approximation ratio that can be achieved by a deterministic local algorithm \cite{suomela13survey}. In many cases, we can apply a straightforward two-step procedure to derive tight lower bounds:
\begin{enumerate}[noitemsep]
    \item Prove tight bounds for anonymous networks (without unique identifiers).
    \item Apply a simulation argument \cite{goos13local-approximation} to show that unique identifiers do not help.
\end{enumerate}
However, there are some isolated examples of natural questions in which the above two-step procedure fails badly. Perhaps the most intriguing example is \emph{dominating sets on planar graphs}:
\begin{enumerate}[noitemsep]
    \item We do not have tight bounds for this problem in anonymous networks.
    \item Planar graphs are not closed under lifts, and therefore the simulation argument~\cite{goos13local-approximation} cannot be applied.
\end{enumerate}
In this work we are interested in the smallest $\alpha$ such that there is a deterministic local algorithm that finds an $\alpha$-approximation of a minimum dominating set in any planar graph. The current bounds are very far from being tight:
\begin{itemize}[noitemsep]
    \item $5-\epsilon < \alpha \le 636$ for anonymous networks \cite{czygrinow08fast,wawrzyniak13mds-brief},
    \item $5-\epsilon < \alpha \le 52$ in the $\LOCAL$ model \cite{czygrinow08fast,lenzen11phd,lenzen13mds,wawrzyniak14mds-analysis}.
\end{itemize}
In this work we give the first improvement on the lower bounds in six years: we prove a lower bound $\alpha > 7-\epsilon$ for both models, for any positive constant $\epsilon$.

\section{Proof Overview}

Let $\A$ be a deterministic distributed algorithm with running time $T = O(1)$ in the $\LOCAL$ model. Assume that $\A$ finds a dominating set $D = \A(G)$ in any planar graph~$G$.

\begin{SCfigure}
\includegraphics[page=1]{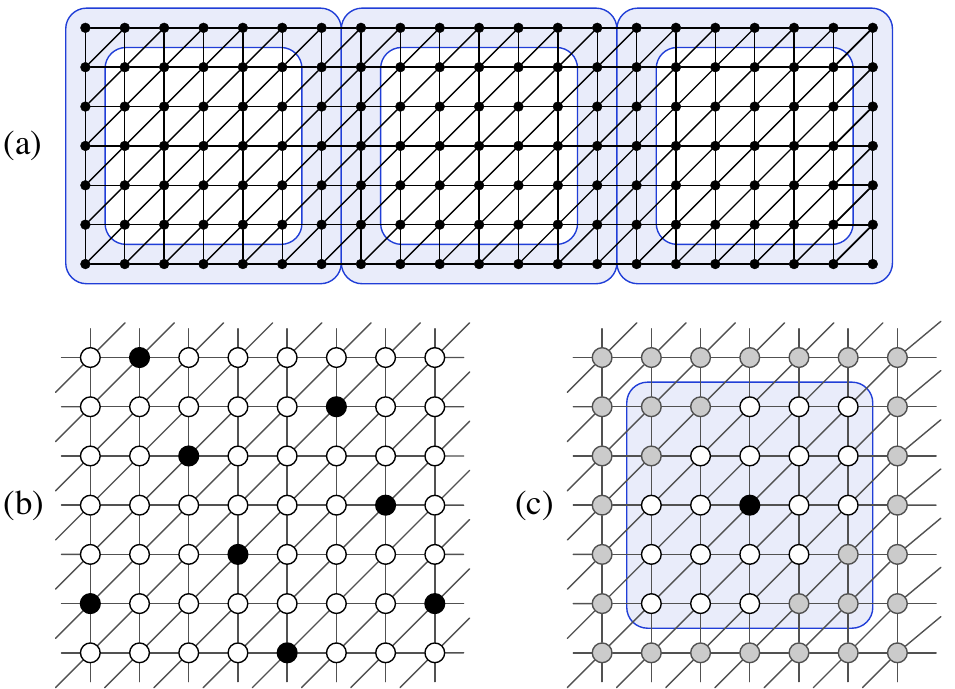}
\caption{(a)~Construction of graph $G$ for $T = 1$, $m = 7$, and $r = 3$. There are $3$ blocks. In each block there are $7 \times 7$ nodes: $5 \times 5$ internal nodes (white area), surrounded by a boundary area of width $1$ (shaded). (b)~A dominating set $D^*$ of $G$ that contains only a fraction $1/7$ of internal nodes. (c)~The local output of an internal node $v$ (black node) only depends on its radius-$T$ neighbourhood (white nodes, here $T=2$). In particular, if we know the unique identifiers in the $k \times k$ region $R_v$ around $v$ (shaded area), we know the local output of node~$v$.}\label{fig:G}
\end{SCfigure}

Pick sufficiently large $m \gg T$ and $r$. Let $m' = m-2T$. We will construct a planar graph $G$ with $n = m^2 r$ nodes as shown in Figure~\ref{fig:G}a. There are $r$ \emph{blocks} with $m \times m$ nodes in each block. The nodes of each block are partitioned to \emph{internal nodes} and \emph{boundary nodes}: there are $m' \times m'$ internal nodes, and they are surrounded by boundary areas of width $T$. Let $B_i$ be the set of nodes in block $i$, and let $I_i \subseteq B_i$ be the set of internal nodes in block $B_i$. We will prove the following lemma.
\begin{lemma}\label{lem:main}
    For any $m$ and any sufficiently large $r$, we can assign unique identifiers in $G$ so that $I_i \subseteq \A(G)$ for all $1, 2, \dotsc, r-\ell$, for some $\ell = o(r)$.
\end{lemma}
In other words, all internal nodes of blocks $1, 2, \dotsc, r-\ell$ are in the dominating set $D = \A(G)$ produced by algorithm $\A$. Now if we choose large enough $m$ and $r$, we can make the contributions of the boundary nodes and the contributions of the remaining $o(r)$ blocks arbitrarily small. In particular, for any positive constant $\epsilon'$, we can pick $m$ and $r$ such that $|D| \ge (1-\epsilon') n$.

On the other hand, there is a dominating set $D^*$ which contains only a fraction $1/7$ of the internal nodes; see Figure~\ref{fig:G}b. Therefore $|D^*| \le (1/7 + \epsilon') n$, and the claim follows: for any positive constant $\epsilon$ we can show that algorithm $\A$ cannot find a factor $7-\epsilon$ approximation of a minimum dominating set on planar graphs.

\section{Proof of Lemma~\ref{lem:main}}

The proof uses the strategy of repeated applications of Ramsey's theorem; cf.\ Czygrinow et al.~\cite[Lemma~4]{czygrinow08fast}. We will use the notation $\A(G,v) \in \{0,1\}$ to refer the \emph{local output} of node $v$ when we apply algorithm $\A$ to graph $G$; we have $\A(G,v) = 1$ if node $v$ is in the dominating set computed by algorithm~$\A$. By definition, $\A(G,v)$ only depends on the radius-$T$ neighbourhood of $v$ in $G$.

Let $k = 2T+1$, $K = k^2$, and $M = m^2$. Consider any internal node $v \in I_i$ of any block $B_i$. The structure of graph~$G$ in the radius-$T$ neighbourhood does not depend on the choice of $v$. Hence the local output of node $v$ only depends on the unique identifiers in the local neighbourhood. The local neighbourhood is contained within a rectangular $k \times k$ region $R_v \subseteq B_i$; see Figure~\ref{fig:G}c.

Let $V = \{1,2,\dotsc,n\}$ be the set of unique identifiers. Consider any $K$-subset of identifiers $X \subseteq V$, $|X| = K$. We will associate a \emph{colour} $c(X) \in \{0,1\}$ with each such set, as follows:
\begin{enumerate}[noitemsep]
    \item Pick an internal node $v$.
    \item Assign the identifiers from $X$ to region $R_v$ in an increasing order by rows: the smallest $k$ identifiers to the bottom row from left to right, etc. Assign the identifiers from $V \setminus X$ to the remaining nodes arbitrarily.
    \item Apply algorithm $\A$, and set $c(X) = \A(G,v)$.
\end{enumerate}

Now we have defined a colouring of all $K$-subsets of $V$; by restriction, we also have a colouring of all $K$-subsets of any $V' \subseteq V$. We say that $Y \subseteq V$ is \emph{monochromatic} if $c(X_1) = c(X_2)$ for any $K$-subsets $X_1$ and $X_2$ of $Y$. By Ramsey's theorem \cite{ramsey30problem} there exists an integer $N = N(K,M)$ such that the following holds: if $V'$ is any $N$-subset of $V$, then there always exists a monochromatic subset $Y \subseteq V'$ of size $M$.

Now we will pick $r$ and $\ell$ so that $\ell M > N$ and $\ell = o(r)$. Let $V_1 = V$. For each $i = 1, 2, \dotsc, r-\ell$, we define the identifiers of block $i$ as follows.
\begin{enumerate}[noitemsep]
    \item As $|V_i| \ge N$, we can find a monochromatic subset $Y_i \subseteq V_i$ of size $M$.
    \item Assign the identifiers from $Y_i$ to block $B_i$ in an increasing order by rows: the smallest $m$ identifiers to the bottom row from left to right, etc.
    \item Set $V_{i+1} = V_i \setminus Y_i$.
\end{enumerate}
Finally, assign the remaining $\ell M$ identifiers from $V_{r-\ell+1}$ to blocks $r-\ell+1, \dotsc, r$ arbitrarily.

To complete the proof, consider a block $i$, where $1 \le i \le r-\ell$. Let $v \in I_i$ be an internal node of the block. Consider the $k \times k$ region $R_v$ around $v$, and let $X_v$ be the set of unique identifiers assigned to region $R_v$. Observe that the identifiers of $X_v$ are assigned in an increasing order by rows. It follows that $\A(G,v) = c(X_v)$, i.e., the local output of the internal node $v$ is simply the colour of subset $X_v$. Furthermore, $X_v \subseteq Y_i$ and $Y_i$ was monochromatic. Hence all internal nodes of block $i$ produce the same output. The common output cannot be $0$; otherwise there would be nodes that are not dominated. Hence $I_i \subseteq \A(G)$.

\paragraph{Acknowledgements.}

Many thanks to Wojciech Wawrzyniak for discussions. This work was supported in part by the Deutsche Forschungsgemeinschaft (DFG, reference number Le 3107/1-1), by the Academy of Finland, Grant 252018, and by the Research Funds of the University of Helsinki.

\def\UrlFont{\sf\footnotesize}
\setlength{\bibsep}{3pt}
\bibliographystyle{abbrvnat}
\bibliography{planar-mds}

\end{document}